\documentclass{aastex631}

\usepackage{amsmath}

\usepackage[normalem]{ulem}
\usepackage{enumitem}

\usepackage{blkarray}
\usepackage{mathtools}
\usepackage{cancel}

\usepackage{newunicodechar,graphicx} % to get the hawaiian okina
\DeclareRobustCommand{\okina}{%
  \raisebox{\dimexpr\fontcharht\font`A-\height}{%
    \scalebox{0.8}{`}%
  }%
}
\newunicodechar{ʻ}{\okina}

\let\orgautoref\autoref
\providecommand{\Autoref}[1]{\def\equationautorefname{Equations}\orgautoref{#1}}

\providecommand{\Autoreff}[2]{\def\equationautorefname{Equations}\orgautoref{#1} and \ref{#2}}

\renewcommand{\autoref}[1]{\def\equationautorefname{Equation}\orgautoref{#1}}

\newcommand*{\myeqref}[2][Eq.~]{%
  \hyperref[{#2}]{#1(\ref*{#2})}%
}
\def\equationautorefname#1#2\null{%
  Eq.#1(#2\null)%
}

\renewcommand{\leq}{\leqslant}  
\renewcommand{\geq}{\geqslant}  
\newcommand{\pow}[1]{^{#1}}
\newcommand{\poww}[1]{\pow{2}}
\newcommand{\powww}[1]{^{3}}

\newcommand{\pderiv}[2]{\partial_#2 #1}

\newcommand{\bs}[1]{\boldsymbol{#1}}
\newcommand{\mathbfit}[1]{\bs{\mathit{#1}}}
\renewcommand{\a}{{\mathbfit{a}}}
\renewcommand{\b}{{\mathbfit{b}}}
\newcommand{\e}{\hat{\bs{e}}}
\newcommand{\g}{\mathbfit{g}}
\renewcommand{\k}{\mathbfit{k}}

\newcommand{\m}{\mathbfit{m}}

\renewcommand{\v}{{\mathbfit{v}}}
\newcommand{\x}{\mathbfit{x}}
\newcommand{\B}{{\mathbfit{B}}}

\newcommand{\E}{{\mathbfit{E}}}
\newcommand{\F}{{{{}_2F_3}}}
\newcommand{\X}{\mathbfit{X}}
\renewcommand{\P}{{\mathbfit{P}}}
\newcommand{\U}{\mathbfit{U}}

\newcommand{\D}{\mbox{\boldmath$\mathsf{D}$}}
\newcommand{\R}{\mbox{\boldmath$\mathsf{R}$}}
\newcommand{\Q}{\mbox{\boldmath$\mathsf{Q}$}}
\newcommand{\J}{\mbox{\boldmath$\mathsf{J}$}}
\newcommand{\Y}{\mbox{\boldmath$\mathsf{Y}$}}

\newcommand{\vdot}{{\mathbf{\cdot}}}

\newcommand{\grad}{\mbox{\boldmath$\nabla$}}

\renewcommand\div{\grad\vdot}

\newcommand{\alf}{Alfv{\'e}n}

\renewcommand{\a}{{\mathbfit{a}}}
\renewcommand{\b}{{\mathbfit{b}}}
\renewcommand{\k}{\mathbfit{k}}
\renewcommand{\v}{{\mathbfit{v}}}
\renewcommand{\P}{{\mathbfit{P}}}

\newcommand{\lrp}[1]{\left( #1 \right)}
\newcommand{\lrb}[1]{\left[ #1 \right]}

\usepackage{xcolor}
\definecolor{forestgreen(web)}{rgb}{0.13, 0.55, 0.13}

\begin{document}

\title{Exact Nonlinear Decomposition of Ideal-MHD Waves Using Eigenenergies II: Fully Analytical EEDM Equations and Pseudo-Advective Energies\footnote{Released on September, 18th, 2024}}

\author[0000-0002-6408-1829]{Abbas Raboonik}
\affiliation{The University of Newcastle,
University Dr, Callaghan,
NSW 2308, Australia}

\author[0000-0002-1089-9270]{David I. Pontin}
\affiliation{The University of Newcastle,
University Dr, Callaghan,
NSW 2308, Australia}

\author[0000-0002-8259-8303]{Lucas A. Tarr}
\affiliation{National Solar Observatory, 22 Ohi\okina{}a Ku St, Makawao, HI, 96768}



\begin{abstract}
Physical insight into plasma evolution in the magnetohydrodynamic (MHD) limit can be revealed by decomposing the evolution according to the characteristic modes of the system. In this paper we explore aspects of the eigenenergy decomposition method (EEDM) introduced in an earlier study (Raboonik et al. 2024, ApJ, 967:80). The EEDM provides an exact decomposition of  nonlinear MHD disturbances into their component eigenenergies associated with the slow, Alfv\'en, and fast eigenmodes, together with  two zero-frequency eigenmodes. 
Here we refine the EEDM by presenting globally analytical expressions for the  eigenenergies. We also explore the nature of the zero-frequency ``pseudo-advective modes''  in detail. We show that in evolutions with pure advection of magnetic and thermal energy (without propagating waves) a part of the energy is carried by the pseudo-advective modes.
Exact expressions for the error terms associated with these modes -- commonly encountered in numerical simulations -- are also introduced. 
The new EEDM equations provide a robust tool for the exact and  unique decomposition of nonlinear disturbances governed by homogeneous quasi-linear partial differential equations, even in the presence of local or global degeneracies.
\end{abstract}

%,Classical Novae (251) --- Ultraviolet astronomy(1736) --- History of astronomy(1868) --- Interdisciplinary astronomy(804)}


\section{Introduction}
Many astrophysical plasmas can be modelled using the equations of magneto\-hydro\-dynamics (MHD), which support waves that transport energy through the system. The \cite{Alfven1942} and slow and fast magneto\-acoustic waves \citep{banos1955} propagate aniso\-tropically with their nature being dependent on the local state of the plasma.
In a previous paper \citep{Raboonik2024} (henceforth referred to as Paper I) we developed a formalism for analysing a dynamic MHD state in terms of the component waves by ascribing energy densities to the wave modes using the MHD eigensystem. In this paper we extend that analysis and refine its physical interpretation, as described below.
Paper I commenced with a detailed motivation and literature review, which we only summarise here.

Understanding the behaviour of MHD waves in the solar corona has two principal motivations. First, damping of MHD waves by various mechanisms has long been invoked to explain the heating of the chromosphere and corona \citep[e.g.][]{vandoorsselaere2020}. Second, such waves can be used to estimate unknown physical variables via coronal seismology \citep[e.g.][]{demoortel2012,nakariakov2020}. In both applications it is highly desirable to be able to understand the characteristics of propagating disturbances in the context of simple, linear wave modes, whose properties are well studied.   

The correct characterisation of propagating disturbances in MHD simulations is critical when comparing with observations. In the simplest approach, interesting dynamical behaviors are identified in simulations, and are interpreted in the context of linear MHD modes \citep[e.g.][]{fuentes2012,thurgood2017,wyper2022}. 
A more sophisticated approach involves a \emph{decomposition} of an identified propagating disturbance into slow, fast, and Alfv\'en(ic) contributions. One approach often taken is to analyse components of the velocity field parallel and perpendicular to the local magnetic field
\citep[e.g.][]{RosBogCar02aa,2019ApJ...883..179K,yadav2022}, 
and then to make analogies to the MHD modes in the limits $\beta=0$ or $\beta\to\infty$. By contast,
\citet{Tarr2017} identified the characteristic properties of propagating disturbances in an MHD simulation by assessing the balance of magnetic and acoustic energy densities   (fast (slow) waves having a magnetically (acoustically) dominated wave energy for $\beta\ll 1$, and vice versa for $\beta\gg 1$).
This approach utilizes the entire MHD state vector to identify waves instead of just the velocity components and more easily reveals the mixed-mode nature of the waves in regions of moderate $\beta$ that are difficult to interpret in velocity-only analyses. \cite{Raboonik2019} and \cite{Raboonik2021} used a combination of the purely acoustic and magnetic parts of the energy along with the dispersion diagrams of the linear system to trace the wave-modes and locate sites of mode-conversion in 2.5D Hall-MHD.

All of the above methods are inexact. An exact treatment can be made by employing the method of characteristics \citep{Jeffrey1964} based on the MHD eigensystem, which forms the basis of many numerical MHD methods \citep{HarLax83,roe1996}. A linear version has been used to identify waves in solar wind observations \citep{Zank2023}.
Recently, \citet{Tarr2024} demonstrated how the characteristic description of MHD could be used to locally decompose a fully nonlinear MHD system into its  characteristic modes, with a focus on implementation of boundary conditions in simulations.

In  Paper I we provided a mathematical formulation for exact nonlinear energy partitioning of ideal MHD disturbances into their component eigenmodes, dubbed the Eigenenergy Decomposition Method (EEDM). The method was based on the eigendecomposition of the flux matrices $M_q$, which made up the quasi-linear form of the ideal MHD equations given by 
\begin{equation}\label{eq:mhd}
    \partial_t \bs{P} + \sum_{q\in\lrp{x,y,z}} M_q \partial_q \bs{P} = 0,
\end{equation}
where $\bs{P} = \lrp{\rho, v_x, v_y, v_z, B_x, B_y, B_z, p}^T$ is the plasma state (or solution) vector, in which $B_q$ and $v_q$ are, respectively, the $q$-directed components of the magnetic field and plasma velocity, $p$ is the pressure, and $\rho$ is the density. The decomposition is acomplished by diagonalizing the flux matrices as $M_q=R_q\Lambda_qL_q$, where $R_q$ and $L_q$ are the right and left eigenmatrices and $\Lambda_q$ is the diagonal matrix of eigenvalues. The total energy of the plasma is given by $E_\text{tot}=\frac{1}{2}\rho v^2 + p (\gamma -1)^{-1}+B^2(2\mu_0)^{-1}$, and its time rate of change by $\partial_t{E_{\rm{tot}}} = \frac{1}{2} \rho_t v^2 + \rho \bs{v}_t\vdot\bs{v} + \frac{p_t}{\gamma - 1} + \frac{1}{\mu_0} \bs{B}_t\vdot\bs{B} = \sum_{q,m} \partial_t E_{m,q}.$  
The last expression is constructed by inserting the mode-decomposed representation of the time derivative of each primitive variable into $\partial_tE_\text{tot}$ to derive the rate of change in terms of the eigendecomposed modes, and it is the focus of the present study.

The essence of the EEDM was captured by the Equations (10) in Paper I (henceforth referred to as the original EEDM equations), describing the rate of change of the energy density components carried by the five possible MHD modes in the three spatial dimensions $q\in\lrp{x,y,z}$ and two characteristic directions (reverse ($-$) and forward ($+$)). These were the divergence ($m = 1$) and entropy ($m = 2$) \emph{pseudo-advective} (PA) eigenmodes, and the familiar reverse and forward \alf{} ($m = 3,4$), slow ($m = 5,6$), and fast ($m = 7,8$) eigenmodes (where $m$ denotes the mode number). Notice the revised terminology of the first two modes where we suffixed the hitherto pseudo with the phrase ``advective''.
Below, we give new globally analytical expressions for the rates of change of energy in each of the modes. Throughout this paper, we will use the phrases `mode' and `eigenmode' interchangeably, since all the permissible modes of \autoref{eq:mhd} are actually the eigenmodes associated with the flux matrices.

One of the main goals of this paper is to shed more light on the interpretation of the PA modes in terms of the advection of energy densities in the system. This extends and clarifies the analysis of Paper I (especially the results of the Section 4 therein) in which we only substantively discussed the possible numerical ``pitfalls'' (errors) that may contribute (unphysical) energy to these modes, while only briefly and conjecturally touching on their (physically valid) advective nature.
To this end, we devote \autoref{sec:advection} to a more rigorous study of the PA modes using exact advective solutions. In short, contrary to the \alf, slow, and fast modes which can both \emph{propagate}\footnote{Read transport energy along the MHD characteristic curves at characteristic speeds.} and advect\footnote{Read transport energy along the direction of the bulk fluid motion at the rate of the plasma velocity.} energy, the PA modes may only contribute to the transport of energy via advection, and hence the terminology. More precisely, the divergence PA mode (which is purely magnetic) describes the portion of the magnetic energy density that can only be advected by the fluid, while the entropy PA mode (a purely thermodynamic property of the plasma) describes portions of the kinetic and internal energy densities. These are direct consequences of the underlying nature of the ideal-MHD model encapsulated in \autoref{eq:mhd} dictating that each fluid cell evolves such that (i) the volumetric magnetic field always remains divergence-free (which is a fundamental feature of all magnetic fields), and that (ii) the volumetric internal energy changes adiabatically (which is merely an imposed but modifiable assumption). The divergence PA mode has been extensively studied in the context of solar wind \citep[see][and references therein]{Zank2023}. In 2D, it is known as the ``magnetic island mode'', while in 3D it is referred to as the ``flux rope mode''.

Another main objective here is to mathematically improve the behaviour of the EEDM equations by eliminating their singularities using an alternate fully analytical 
set of eigensystems 
based on \cite{roe1996}. In their current form in Paper I, it can be shown that although the time variations of the total wave-energy density remains analytical everywhere, due to the non-associative property of limits, some of the slow and \alf{} component energies can become singular either locally or globally depending on the nature of the solution vector $\bs{P}$ in certain limits where one or more components of the magnetic field approach zero. This has to do with the fact that while the flux matrices are globally well-behaved, their eigensystems (which are the building blocks of the EEDM) need not necessarily be, and may contain singularities depending on their mathematical construction. As will become evident, this improvement is of central importance since the class of such singular (numerical or exact) solutions is not negligible. The updated version of the EEDM equations, now void of any singularities, will be presented in \autoref{sec:EEDMNew}.
Lastly, in \autoref{sec:caveats} we will provide more formal definitions of the above-mentioned numerical pitfalls associated with the PA modes. 

\section{Fully analytical Form of EEDM equations}\label{sec:EEDMNew}
As mentioned before, the core of the EEDM is based on the local eigendecomposition of the flux matrices $M_q$ at each point in space. It is the eigensystems that contain the necessary information to carry out local transformations of the gradient state vector $\partial_q\bs{P}$ onto the local characteristic curves where the MHD modes are distinguishable, and hence decomposable. 
However, care must be taken as although $M_q$ are fully analytical\footnote{Note that by analytical we mean that all the matrix entries are finite within the plasma.} and diagonalizable (or non-defective and hence eigen-decomposable) everywhere in the plasma\footnote{The flux matrices are non-defective even when $\B = \bs{v} = 0$ (resulting in the repeated eigenvalue $\lambda = 0$ of both algebraic and geometric multiplicities of six), which is simply the hydro\-static case.}, their eigensystems may contain singular matrix elements in certain limits, giving rise to singular terms in the EEDM equations. However, such (removable) singularities may be avoided by a judicious choice of normalization of the eigenvectors \citep{brio1988, roe1996}. 

The left $L_q$ (right $R_q$) eigenvectors listed in the Appendix A of Paper I used to derive the original EEDM equations fall into this category. They contain singularities in the rows (columns) associated with the slow (rows (columns) 5 and 6 in $L_q$ ($R_q$)) and \alf{} (rows (columns) 3 and 4 in $L_q$ ($R_q$)). The singular points $\bs{x}_0$ are located wherever $a_q = 0$, $a_{\perp q} = 0$, or (at magnetic nulls where) $a = 0$, in which $a = B/\sqrt{\mu_0 \rho}$ denotes the familiar \alf{} speed with $\mu_0$ being the permeability of vacuum, and the subscript $\perp q$ represents directions perpendicular to $q$.
Additionally, recall that the fast (f) and slow (s) characteristic speeds are given by $c_{\text{f/s},q} = \lrp{a^2 + c^2 \pm \sqrt{a^4 + c^4 + 2 c^2(a^2 - 2a_q^2)}}^{1/2}/\sqrt{2}$, where $c =  \sqrt{\gamma p / \rho}$ is the adiabatic sound speed, with $\gamma$ the heat capacity ratio. Thus, we find that $c_{\text{s},q} = 0$ wherever $a_q = 0$. Consequently, the original mathematical forms of $L_q$ and $R_q$ do not satisfy their respective eigenequations in such limits, i.e., $\lim\limits_{\bs{x}\rightarrow \bs{x}_0}R_q \Lambda_q L_q \neq \lim\limits_{\bs{x}\rightarrow \bs{x}_0}M_q$.

In order to remove such singularities from the EEDM, we adopt the eigensystems provided by \cite{roe1996}, and augment them with an additional (first) row/column to account for the divergence PA mode,  and rederive a new set of EEDM equations. The new left eigenvectors denoted by $\mathbb{L}_{q}$ are given in \autoref{sec:AppendixA}. Subsequently, their associated right eigenvectors can be computed by taking the inverse according to $\mathbb{R}_{q} = \mathbb{L}_{q}^{-1}$, and using some identities discussed therein. In short, these identities concern the dimensionless normalization functions originally proposed by \cite{brio1988} (and modified slightly here for a more compact representation) to remove the eigensystem singularities, defined by
\begin{subequations}\label{eq:alphaBeta}
    \begin{equation}\label{eq:alpha}
        \alpha_{\text{s/f},q} = \lrp{\mathcal{S}_\text{s/f}\frac{c_{\text{f/s},q}^2 - c^2}{c_{\text{f},q}^2 - c_{\text{s},q}^2}}^{1/2},
    \end{equation}
    \begin{equation}\label{eq:beta}
        \beta_{q' \perp q} = \begin{cases} 
      a_{q'}/a_{\perp q} & a_{\perp q} \neq 0 \\
      \frac{1}{\sqrt{2}} & a_{\perp q} = 0  \text{ or }  a = 0 
   \end{cases},
    \end{equation}
\end{subequations}
where $\mathcal{S}_\text{s} = 1$, $\mathcal{S}_\text{f} = -1$, and the subscript $q' \perp q$ means in the direction of $q'$ and perpendicular to $q$ (e.g., $\beta_{y \perp x} = a_y/\sqrt{a_y^2 + a_z^2}$). Note that these are well-behaved functions everywhere; however, the piece-wise definition of $\beta$ enforces a piece-wise form for the new EEDM equations.

Ultimately, based on the new eigensystems, the set of fully analytical EEDM equations are as follows
\begin{subequations}\label{eq:EEDMNew}  
    \begin{equation}\label{eq:div}
       \partial_t{E_{\text{div},q}} = -\frac{B_q \partial_q{B_q}}{\mu_0} v_q,
    \end{equation}
    
    \begin{equation}\label{eq:ent}
       \partial_t{E_{\text{ent},q}} = -\frac{v^2 \lrp{c^2 \partial_q{\rho} - \partial_q{p}}}{2 c^2} v_q,
    \end{equation}

    \begin{equation}\label{eq:alfNew}
        \partial_t{E_{\text{A},q}^{\mp}} = - \big(\sqrt{\mu_0 \rho}\bs{v}\bs{\times}\bs{\mathcal{B}}_q\big)_q  \Big(\big(\sqrt{\mu_0 \rho}\partial_q{\bs{v}} \pm s_{q}\partial_q{\bs{B} }\big)\bs{\times}\bs{\mathcal{B}}_q\Big)_q \dfrac{\lrp{v_q \mp \|a_q\|}}{2 \mu_0}
    \end{equation}

    \begin{eqnarray}\label{eq:sloFasNew}\begin{array}{c}{{\rm\partial }}_{t}{E}_{{\rm{s/f}},q}^{\mp }=\displaystyle  
-\lrp{\mathcal{S}_\text{s/f} \alpha_{\text{f/s},q} \Big(\pm s_q c_{\text{f/s},q} \sqrt{{\mu }_{0}\rho } \bs{v} + c \bs{B}\Big)\vdot\bs{\mathcal{B}}_q + \sqrt{{\mu }_{0}\rho }\alpha_{\text{s/f},q}\Big(\pm c_{\text{s/f},q} v_q - \frac{c^2}{\gamma - 1} - \frac{1}{2} v^2\Big) }\\ \hspace{2.25cm}\times \,\lrp{\mathcal{S}_\text{s/f}\alpha_{\text{f/s},q}  \Big(\pm s_q c_\text{f/s} \sqrt{\mu_0 \rho} \partial_q \bs{v} + c \partial_q \bs{B}\Big)\vdot \bs{\mathcal{B}}_q +\sqrt{\mu_0 \rho}\alpha_{\text{s/f},q}\Big(\pm c_{\text{s/f},q}\partial_q v_q - \dfrac{1}{\rho} \partial_q p\Big)}\dfrac{ \lrp{v_q \mp c_{\text{s/f},q}}}{2 \mu_0{c}^{2}}\end{array} 
    \end{eqnarray}
\end{subequations}
wherein 
$\bs{\mathcal{B}}_q = \lrp{\beta_{x\perp q}\lrp{1-\delta_{q,x}},\beta_{y\perp q}\lrp{1-\delta_{q,y}},\beta_{z\perp q}\lrp{1-\delta_{q,z}}}$ with $\delta_{q,q'}$ representing the Kronecker delta function, $E$ denotes the energy density, the subscripts div and ent refer to the divergence and entropy PA modes, and A, s, and f stand for the \alf, slow, and fast modes, and $s_q = \text{sgn}(a_q)$.  
Note that \autoref{eq:sloFasNew} defines four modes through all four combinations of $\pm$ and the subscripts s/f. The use of $\bs{\mathcal{B}}_q$ gives a particularly compact design, but the meanings are quite simple in that all it does (aside from accounting for the piece-wise definition of $\beta_{q'\perp q}$) is select the components that are perpendicular to $q$ as follows. For an arbitrary vector $\bs{d}: \lrp{\bs{d}\bs{\times}\bs{\mathcal{B}}_q}_q = (\bs{d}_\perp \bs{\times} \hat{\bs{b}}_\perp)_q$ and $\bs{d} \vdot \bs{\mathcal{B}}_q = \bs{d}_\perp \vdot \hat{\bs{b}}_\perp$, where $\hat{\bs{b}}_\perp$ is either the unit vector of the magnetic field that is perpendicular to the direction $q$ if $\|\bs{\B}_\perp\| > 0$, or $\hat{\bs{b}}_\perp = (1-\delta_{q,x},1-\delta_{q,y},1-\delta_{q,z})/\sqrt{2}$ if $\|\bs{\B}_\perp\| = 0$.
Thus, equipped with the piece-wise variable $\bs{\mathcal{B}}_{q}\lrp{\beta_{q'\perp q}}$, \Autoref{eq:EEDMNew} provide the fully analytical EEDM equations.

Note that the PA energy components given by \Autoreff{eq:div}{eq:ent} are identical to those of Paper I.
This is expected as their original associated rows (columns) in $L_q$ ($R_q$) were fully analytical, and hence have been carried forward (aside from a factor of $\rho c^2$) in $\mathbb{L}_q$ ($\mathbb{R}_q$). We remind the reader that, mathematically speaking, the two PA modes are zero-frequency degenerate eigen-modes of the ideal-MHD system on account of their identical eigenvalues $v_q$ (algebraic multiplicity of two), but distinguishable due to their linearly independent eigenvectors (geometric multiplicity of two also, and hence non-defective). In fact, more generally, since the EEDM takes advantage of the \emph{globally} non-defective nature of the flux matrices, and uses the resulting linearly independent eigenvectors as distinct mode identifiers, it provides a powerful decomposition tool that remains valid even in the presence of local/global mode degeneracies.

Additionally, the reader may gain more insight into the similarities and differences between the \alf{} and magneto\-acoustic branches by checking that since  $\lrp{\bs{B}\bs{\times}\bs{\mathcal{B}}_q}_{q} = 0$, we could have rewritten \autoref{eq:alfNew} with this term (times $s_q$) added to the first bracket on the right-hand-side (RHS), and hence constructing a combination of terms resembling the ones involving $\mathcal{\B}_q$ in \autoref{eq:sloFasNew}. Doing so, we may then compare the two equations and see that the \alf{} branches are made purely out of the cross products of $\bs{\mathcal{B}}_q$, and $\bs{v}$ and $\bs{B}$ and their derivatives, while the purely magnetic parts (the first terms in each line of \autoref{eq:sloFasNew}) of the slow/fast branches involve the dot product of these vectors. This is reminiscent of the incompressive vortical polarization of the (torsional) \alf{} wave, and the compressive linearly polarized (kink and sausage) nature of the magneto\-acoustic waves.

Finally, using \Autoref{eq:EEDMNew} and assigning a mode number $m\in[1..8]$ to each equation, the individual eigenenergy densities can be obtained from
\begin{equation}
    E_{m,q} = \int_{t_0}^{t}{\partial_t E_{m,q}}\,dt', \quad \text{with }E_{m,q}\bigg|_{t = t_0} = 0,
\end{equation}
satisfying $E_\text{tot} - E_0 = \sum_{q\in\lrp{x,y,z}}\sum_{m=1}^8 E_{m,q}$, in which $E_0$ denotes the total energy at $t = t_0$. Note that $E_0$ may describe an already \emph{dynamic} state.

\section{Interpretation of pseudo-advective modes: advective solutions}\label{sec:advection}
In this section, we address the physical interpretation of the pseudo-advective modes in the EEDM. We do this by formulating two exact, nonlinear MHD solutions that describe pure advection.

\subsection{1.5D Straight Flux Tube Advection}\label{subsec:1Dflux}
Consider an untwisted magnetic flux tube with a Gaussian cross-section
parallel to the $z$ axis, moving at a constant rate $v_0$ in the $x$ direction. 
We may construct the exact MHD solution describing such a system as follows:
\begin{subequations}
\begin{equation}\label{eq:exact1.5D}
    \bs{P}_\text{adv} = \lrp{\rho, v_0, 0, 0, 0, 0, B_0\lrp{1 + G}, p_0 - \frac{B_0^2(1+G)^2}{2 \mu_0}},
\end{equation}
\begin{equation}
        G(x,y,t) = \exp\lrp{-\frac{\lrp{x - v_0 t}^2 + y^2}{2 \sigma^2}},
\end{equation}
\end{subequations}
in which $p_0$ (assumed sufficiently large so that $p>0$) and $B_0$ are constant, $\rho$ is an arbitrary function of $x - v_0 t$ and $y$, $G$ is the Gaussian function, and $\sigma$ sets the width of the flux tube in the $xy$ plane.
Note that this solution reduces the MHD equations to
\begin{equation}\label{eq:system1.5}
    \begin{cases}
        \partial_t P_n + v_0 \partial_x P_n = 0 & n \in \left[1,7,8\right] \\
        \partial_t P_n = 0 & \text{otherwise}
    \end{cases},
\end{equation}
where  $P_n$ with $n \in \left[1..8\right]$ represents the $n$-th primitive variable in the solution vector $\bs{P} = \lrp{\rho, \bs{v}, \B, p}$. The first equation controls how the initial conditions on $\rho$, $B_z$, and $p$ are simply advected, while the second equation maintains the equilibrium (i.e., the force balance between the magnetic and gas pressures) and preserves the initial conditions on the remaining variables. The total energy change for this system is given by
\begin{equation}\label{eq:energy1.5}
    \partial_t E_\text{tot} = -\frac{1}{2} v_0^3 \partial_x \rho - \frac{\partial_x p}{\gamma - 1} v_0 - \frac{B_z \partial_x B_z}{\mu_0} v_0
\end{equation}

We may now apply \Autoref{eq:EEDMNew} to energy-decompose the above solution. This process can be significantly simplified by noting that all EEDM terms with $q = y, z$ would return zero since the only primitive variables that enter the EEDM equations are the ones that evolve with time (i.e., $\partial_t P_n \neq 0$), which can be seen from \Autoref{eq:system1.5} to only contain gradients in the $x$ direction.
Therefore, the only non-zero components are found to be due to the PA entropy mode and the forward and reverse slow modes along $x$, and are as follows
\begin{subequations}
    \begin{equation}\label{eq:ent1.5}
        \partial_t E_{\text{ent},x} = \frac{v_0^2}{2} \lrp{\frac{\partial_x p}{c^2} - \partial_x\rho} v_0,
    \end{equation}
    
    \begin{equation}\label{eq:slow1.5}
        \partial_t E_{\text{s},x}^\mp = \frac{1}{2}\lrp{- \frac{\partial_x p}{2 c^2} v_0^2 - \frac{\partial_x p}{\gamma -1}- \frac{B_z\partial_x B_z}{\mu_0}} v_0,
    \end{equation}
\end{subequations}
whose net result (summation of all three components) perfectly recovers the time variation of the total energy given by \autoref{eq:energy1.5}. 
Note that for this specific choice of solution we have $a_x = c_{\text{s},x} = 0$, thereby rendering the \alf{} and slow branches into zero-frequency modes, and hence degenerate with the PA modes. Mathematically, this degeneracy is due to the repeated eigenvalue $\lambda_m = v_0 \forall m \in \{\text{ent}, \text{div}, \text{A}^{\mp}, \text{s}^{\mp}\}$ of algebraic and geometric multiplicities of six, i.e., the eigenvectors are still distinct. 

The EEDM detects no energy variations due to the divergence, \alf, or fast modes. Instead, the entirety of the (advected) energy density is carried by a combination of the entropy PA mode,
as anticipated, but also in equal parts (since $\partial_t E_{\text{s},x}^- = \partial_t E_{\text{s},x}^+$) by \emph{non-propagating} reverse and forward slow modes. These modes trivially describe the advection when the slow characteristic speed ($c_{\text{s},q}$) is zero, but as will be seen in the next subsection, the characteristic speeds need not necessarily vanish in order to have non-propagating slow, \alf, or even fast modes. In fact, in spite of the fast speed being strictly positive ($c \leq c_{\text{f},q}$), the superposition of the reverse and forward fast modes may still allow for 
the advection of energy to be described in part by non-propagating fast waves.

Note that in \autoref{eq:ent1.5}, $\rho=\rho(x-v_0t,y,z)$ is arbitrary, and we can distinguish between two cases. First, for an isentropic solution (i.e., in the absence of \emph{Lagrangian} entropy  gradients; $dk/dt = 0$, where $k$ is the polytropic coefficient satisfying $p = k \rho^\gamma$, and $d/dt$ is the Lagrangian time derivative) the entropy PA mode would be identically zero, as expected, and the state would be characterised entirely by the slow mode. Conversely, in the presence of entropy inhomogeneities, a part of the (advected) energy shows up in the PA mode (which may also be non-zero if there is numerical heating -- see section \ref{sec:caveats}). 
Either way, the transport of the equilibrium state via advection is described entirely by the sum of the reverse and forward slow modes.

We emphasise that this example reveals the dual nature of the slow mode in the EEDM: it describes both the propagation of slow waves (provided other solutions characterizing dynamic non-equilibrium states) and a portion of the advected energy in the form of zero-frequency (or non-propagating) modes. However, given the characteristically bidirectional ($\mp$) nature of such modes, the way in which they describe energy advection is different to the PA modes. The statements apply equally well to the fast and \alf{} modes, as we shall see below.

\subsection{2D Twisted Flux Advection}\label{subsec:2Dflux}
Consider this time a twisted magnetic flux tube oriented along the $z$ axis, localized in the $xy$ plane, and advected at a constant rate $\bs{v}_0 = \lrp{v_0,0,0}$. The MHD behaviour of such a system may be given by the following solution vector
\begin{subequations}\label{eq:solution2D}
    \begin{equation}
        \bs{P}_{\text{adv}} = \lrp{\rho, \bs{v}_0, \B, p_0 - U_B - B^2 / 2 \mu_0}
    \end{equation}
    \begin{equation}
        \bs{B} = \lrp{B_1 y G, -B_1\lrp{x - v_0 t}G, B_0},
    \end{equation}
    \begin{equation}
        U_B = -\frac{\sigma^2 G^2 B_1^2}{2 \mu_0},
    \end{equation}
\end{subequations}
in which $B_0$, $B_1$, and $p_0$ are constant, $U_B$ is the magnetic tension potential satisfying $-\nabla U_B = \bs{B}\vdot\nabla\B/\mu_0$. As before, $\rho$ is an arbitrary function of $(x-v_0t,y,z)$ and $G$ is the same Gaussian function \citep[which is stable up to a certain level of twist as per][]{suydam1958}. Note that the magnetic tension force due to $U_B$ is purely radial, and hence works in conjunction with the magnetic pressure to balance out the gas pressure.
Substituting  this solution into the MHD equations we find
\begin{equation}\label{eq:system2}
    \begin{cases}
        \partial_t P_n + v_0 \partial_x P_n = 0 & m \in \lrb{1, 5, 6, 8} \\
        \partial_t P_n = 0 & \text{otherwise}
    \end{cases}.
\end{equation}
From these equations, the total energy time variation is then found to be
\begin{equation}\label{eq:Etot2D}
    \partial_t{E_{\text{tot}}} = - \frac{\partial_x \rho}{2} v_0^3 - \frac{B_x \partial_x B_x}{\mu_0} v_0 - \frac{B_y \partial_x B_y}{\mu_0} v_0 -\frac{\partial_x p}{\gamma -1} v_0.
\end{equation}

Performing the EEDM on $\bs{P}_\text{adv}$ using \Autoref{eq:EEDMNew}, after extensive simplifications, yields
\begin{subequations}
    \begin{equation}
        \partial_t{E_{\text{div},x}} = -\frac{B_x \partial_x{B_x}}{\mu_0}v_0,
    \end{equation}
    \begin{equation}\label{eq:ent2D}
        \partial_t E_{\text{ent},x} = \frac{v_0^2}{2c^2}  \lrp{\partial_x p - c^2 \partial_x\rho} v_0,
    \end{equation}
    \begin{equation}\label{eq:alf+-2D}
        \partial_t{E_{\text{A},y}^\leftrightarrow} =  \frac{a_z^2}{a_{\perp y}^2}\frac{B_y \partial_y{B_x}}{\mu_0}v_0
    \end{equation}

    \begin{equation}\label{eq:s/f2D}
        \partial_t{E_{\text{s/f},x}}^\leftrightarrow = -\frac{1}{c^2}\lrp{\alpha_{\text{f/s},x} c a_{\perp x} - \mathcal{S}_{\text{s/f}}\alpha_{\text{s/f},x}\Big(c_{\text{s/f},x}^2 + \frac{c^2}{\gamma - 1}+ \frac{1}{2}v_x^2\Big)}\lrp{\dfrac{ \alpha_{\text{f/s},x} c}{a_{\perp x}}\dfrac{B_y \partial_x B_y }{\mu_0 } - \mathcal{S}_{\text{s/f}}\alpha_{\text{s/f},x}\partial_x p} v_0
    \end{equation}

    \begin{equation}
        \partial_t{E_{\text{s/f},y}}^\leftrightarrow = \dfrac{\alpha_{\text{f/s},x} a_x a_y}{a_{\perp y} c}\lrp{\dfrac{\alpha_{\text{f/s},x} c}{a_{\perp y}}\dfrac{B_x \partial_y B_x}{\mu_0} - \mathcal{S}_\text{s/f}\alpha_{\text{s/f},x}\partial_y p} v_0
    \end{equation}
\end{subequations}
in which $\leftrightarrow$ denotes the superposition (sum) of both the reverse and forward modes. The reader can check that the sum of all the eigenenergy variations above recovers \autoref{eq:Etot2D}.
Note that even though the individual reverse and forward slow, \alf, and fast modes are apparently non-degenerate due to their non-zero characteristic speeds in this case, their respective superposition ($\leftrightarrow$) reduces to a PA-degenerate form. This is the expected behaviour since the solution vector here is purely advective, and hence all the information has to travel at $v_0$. Thus, this solution confirms that the PA modes 
describe advection of a portion of the energy, and that not only do the slow, \alf, and fast modes 
describe propagating wave modes, but under general circumstances a portion of the advected energy in the system also shows up in those modes. A notable difference between this and the previous example is the appearance of a non-zero energy density variation in the `div' PA mode. This reveals that a part of the magnetic energy density variation is in general captured by this `div' term: specifically, the contribution
related to the variation in the direction of the flow of the field component in that direction.

Moreover, we conjecture that in certain instances one may interpret the phenomenon of the reduction of the superposed reverse and forward modes into non-propagating waves to be equivalent to the \emph{counter-propagation} of the oppositely directed modes within the same branch (e.g., the reverse and forward modes of the slow branch) creating PA-degenerate modes. We base this speculation on our observation of the reverse and forward eigenenergies generating oppositely propagating disturbances in a magnetic twist relaxation simulation. However, the details of this simulation are beyond the scope of this paper and will be explored in a future study.

\section{Quantifying the effects of numerical divergence and dissipation}\label{sec:caveats}
In this section, we will review a few mathematical caveats pertaining to the two PA modes and possible limiting cases of the method that were left out of Paper I.
In deriving the EEDM equations (both here and in Paper I), we eliminated the field-divergence term from the induction equation and assumed that any changes in the internal energy would occur adiabatically. Mathematically, of course $\div{\B} = 0$ should always hold, while the adiabatic assumption would be true if the entropy was constant in the Lagrangian frame, i.e., $\rho^\gamma dk/dt = 0$.
However, when numerically solving the ideal-MHD equations, it is possible that these conditions are not precisely met due to inevitable numerical imprecision. This can result in spurious (unphysical) energies appearing in the PA modes.
We shall here derive the energy error terms associated with possible numerical departures from the zero field-divergence principle and the adiabatic assumption. 

\subsection{Divergence PA Energy Error}
Let us assume that due to inherent numerical errors, $\div{\B} = \Delta_B$, for some non-zero $\Delta_B$. The ideal-MHD induction equation can then be written as 
\begin{equation}\label{eq:3DidealMHD:induction}
    \pderiv{\bs{B}}{t} - \bs{B}\vdot\bs{\nabla}\bs{v} + \bs{B}\div{\bs{v}} + \bs{v}\vdot\bs{\nabla}\bs{B} = \bs{v} \Delta_B.
\end{equation}
The process of deriving the error term due to the numerical generation of field divergence depends on how one interprets the RHS term. Since $\bs{v} \Delta_B$ describes the numerical error in the computation of a primitive variable ($\B$), which can cyclically feed back into the evolution of $\bs{P}$, it can be regarded as an internal fictitious (coupling) effect inherent to the numerical scheme. This can be reflected in the MHD equations through apt modifications to the flux matrices (which would inevitably lead to modifications to the characteristics) as follows
\begin{subequations}\label{eq:modifiedM}
    \begin{equation}
    \bar{M}_x^{i,j} = M_{x}^{i,j}  -\frac{\Delta_B v_x}{\partial_x B_x}  \delta_{i,5} \delta_{j,5}
\end{equation}
\begin{equation}
    \bar{M}_y^{i,j} = M_{y}^{i,j}  -\frac{\Delta_B v_y}{\partial_y B_y}  \delta_{i,6} \delta_{j,6}
\end{equation}
\begin{equation}
    \bar{M}_z^{i,j} = M_{z}^{i,j}  -\frac{\Delta_B v_z}{\partial_z B_z} \delta_{i,7} \delta_{j,7},
\end{equation}
\end{subequations}
which recovers the full set of modified ideal-MHD equations, now accounting for field-divergence errors. Here $M_{q}$ is the original flux matrix with $\Delta_B = 0$. 

The new rectified eigensystems associated with $\bar{M}_q$ now inherit slightly modified characteristic slow/fast speeds due to the added error term. Specifically, we have $\bar{c}_{\text{s/f},q} = -\text{sgn}(\partial_q B_q) c_{\text{s/f},q}$. However, the mathematical form of all the EEDM \Autoref{eq:EEDMNew} remains unaltered, except for \autoref{eq:div} associated with the divergence PA mode, to which the following error term is added on the RHS,
\begin{equation}\label{eq:errorDiv}
    \partial_t \epsilon_{\text{div},q} = \frac{B_q \Delta_B}{\mu_0} v_q.
\end{equation}
Integrating the above with respect to time and summing over $q$ yields the net amount of energy $\epsilon_{\text{div}}$ due exclusively to numerical divergence generation.

Note that the new characteristic speeds  $\bar{c}_{\text{s/f},q}$ only affect the sign of the slow/fast eigenenergy variations given by \autoref{eq:sloFasNew}, leaving the magnitude of their collective contribution to the total energy unaltered. Moreover, one could also argue that since the generation of $\Delta_B$ depends on the specific choice of the numerical scheme as well as the simulation code, the coupling effects due to the RHS of  \autoref{eq:3DidealMHD:induction} could be modelled to represent an external source term, thereby leaving the MHD characteristics entirely untouched. Either way, \autoref{eq:errorDiv} remains valid as an exact measure of the field-divergence energy error. Thus, sufficiently small values of $\Delta_B$ such that $\epsilon_\text{div} \ll E_\text{div}$ would ensure high precision of \autoref{eq:div} in computing the true physically valid amount of advected magnetic energy. 

\subsection{Entropy PA Energy Error}
Now suppose that the adiabatic assumption breaks down on account of numerical heating. Then, the Lagrangian energy equation can be modified as
\begin{equation}
    \frac{\rho^\gamma}{\gamma -1} \frac{d k}{dt} = \partial_t\epsilon_\text{ent},
\end{equation}
where $\epsilon_\text{ent}$ is the amount of energy lost from the total energy due to numerical damping. Using the polytropic equation, we have
\begin{equation}
    \frac{1}{\gamma -1}\rho^\gamma \frac{d}{dt}\lrp{\frac{p}{\rho^\gamma}} = \frac{d p}{dt} - c^2 \frac{d\rho}{dt} = \partial_t\epsilon_\text{ent}.
\end{equation}
Expanding this equation in terms of the Eulerian variables yields
\begin{equation}
    \lrp{\partial_t p - c^2 \partial_t \rho} + \bs{v}\vdot\lrp{\nabla p - c^2 \nabla \rho} =  \lrp{\gamma -1} \partial_t\epsilon_\text{ent}.
\end{equation}
Multiplying both sides of the equation by $v^2/2 c^2$ to construct $\partial_t E_\text{ent}$ out of the second term on the LHS gives
\begin{equation}
    \frac{v^2}{2 c^2}\lrp{\partial_t p - c^2 \partial_t \rho} + \sum_{q}\partial_t E_{\text{ent},q} = \frac{v^2}{2 c^2}\lrp{\gamma -1} \partial_t\epsilon_\text{ent},
\end{equation}
which then can be used to find the energy error term according to
\begin{equation}\label{eq:errorEnt}
    \epsilon_\text{ent} = \frac{1}{\gamma -1}\int{\lrp{\lrp{\partial_t p - c^2 \partial_t \rho} + \frac{2 c^2}{v^2}\sum_{q}\partial_t E_{\text{ent},q}}\,dt}.
\end{equation} 
Similar to the argument made for the divergence PA energy error, the numerical scheme would feasibly uphold the adiabatic assumption if $\epsilon_\text{ent} \ll E_\text{ent}$.

\section{Conclusions}
This paper completes Paper I \citep{Raboonik2024} by introducing the enhanced fully analytical EEDM \Autoref{eq:EEDMNew}, which remain valid even at magnetic null-points (equivalent to the hydro\-dynamic limit), where all but the fast modes become degenerate with the PA modes, 
and the fast modes reduce to sound waves. It also expands on their results by addressing the true physical nature of the ``pseudo-modes'' defined and speculatively interpreted therein, as well as providing exact mathematical expressions associated with the fictitious numerical energies attributed to such modes.

Although valid as an exact mathematical procedure for breaking down the energy contents of nonlinear disturbances governed by homogeneous quasi-linear partial differential equations in terms of their component eigenmodes, Paper I used a set of ideal-MHD eigensystems which contained \emph{removable} singularities wherever one or more components of the magnetic field would tend to zero. While well-behaved elsewhere, this would render the original EEDM equations therein indeterminate for solution vectors containing such limits. Although there are \emph{approximate} numerical ways of treating such singularities, we surmounted the problem entirely and exactly in \autoref{sec:EEDMNew} by adapting an alternative set of fully analytical eigensystems originally set forth by \cite{brio1988} and later extended by \cite{roe1996}. Thus, \Autoref{eq:EEDMNew} present the complete set of globally analytical EEDM equations for non-gravitational ideal-MHD systems. Furthermore, due to the EEDM's use of the MHD eigenvectors as primary eigenmode identifiers,  \Autoref{eq:EEDMNew} remain perfectly valid even in the case of local or global cross-modal degeneracies.

Additionally, here we established an unequivocal physical interpretation of what was previously defined as (the divergence and entropy) ``pseudo-modes'', whose physically valid energy contributions via advection were only conjecturally touched on in Paper I.
Using the two exact purely advective solutions proposed in \autoref{sec:advection}, we achieved this by demonstrating that they do indeed carry a part of the total energy along the flow via strict advection.
This imposed a modification on the previous terminology to reflect such pure advective nature, and hence they were renamed as the \emph{pseudo-advective} (PA) modes. It was also demonstrated that pure advection is not described solely by the PA modes, but that a part of the advected energy may also appear in the reverse and forward slow, \alf, and fast modes. This was shown to take place in the following two scenarios: (i) wherever $a_q = 0$ (and consequently $c_{\text{s},q} = 0$), and hence rendering the reverse and forward slow and \alf{} modes degenerate with the PA modes (see \autoref{subsec:1Dflux}), and (ii) when the superposition (sum) of the reverse and forward modes of a branch yields a PA-degenerate mode (see \autoref{subsec:2Dflux}). Furthermore, we conjectured that the latter (ii) may be a consequence of the counter-propagating property of the reverse and forward modes, reducing the net result into non-propagating waves. Moreover, given the global equilibrium states of our advective solutions, it is also \emph{suspected} that such nonPA-to-PA degeneracies may be possible only where the restoring forces are in perfect balance. However, further investigation is required to substantiate this suspicion.
Regardless of the details, the most salient point here is that the slow, \alf, and fast modes describe both the corresponding propagating waves, as well as advection of a portion of the energy.

Finally, in addition to the physically valid PA eigenenergies, when applying the EEDM to numerical solutions,
the \emph{numerical pitfalls} extensively discussed in Paper I can also give rise to unphysical energies. To this end, we presented \autoref{sec:caveats} which aimed at evaluating exact expressions capturing such fictitious energies arising from inevitable numerical errors. Thus, \Autoreff{eq:errorDiv}{eq:errorEnt} can be used to determine the numerical fidelity of ideal-MHD solutions as well as the amount of energies erroneously attributed to the PA modes due to the existence of numerical magnetic field divergence and artificial heating.

\section*{acknowledgments}
The authors would like to thank Dr. Sahel Dey for his valuable help in replicating a previous simulation in the Pencil code.
This work was performed on the OzSTAR national facility at Swinburne University of Technology. The OzSTAR program receives funding in part from the Astronomy National Collaborative Research Infrastructure Strategy (NCRIS) allocation provided by the Australian Government, and from the Victorian Higher Education State Investment Fund (VHESIF) provided by the Victorian Government.
Computations were run on the Australian National Computational Infrastructure's Gadi machine through an award from Astronomy Australia Ltd.'s Astronomy Supercomputer Time Allocation Committee.
L.A.T. is supported by the National Solar Observatory.

\software{Analytic work performed in Wolfram Mathematica \citep{Mathematica}. Simulations performed in Lare 3.4.1 \citep{ARBER2001151}. Numerical analysis performed in AutoParallelizePy \citep{Raboonik_AutoParallelizePy_2024}, NumPy \citep{Harris:2020}, and SciPy \citep{Virtanen2020}, and figures were prepared using the Matplotlib library \citep{Hunter:2007}, all in Python3.}





\appendix      

\section{Roe and Balsara Eigenvectors}\label{sec:AppendixA}
The eigenvalues and left eigenvectors associated with the flux matrices $M_q$ in the three Cartesian coordinates, adopted from \cite{roe1996} and augmented to account for the divergence PA mode not considered therein, is given by
\begin{subequations}
    \begin{equation}
    \Lambda_q = \text{diag}\lrp{v_q,v_q,v_q-    \|a_q\|,v_q+\|a_q\|,v_q-c_{\text{s},q},v_q+c_{\text{s},q},v_q-c_{\text{f},q},v_q+c_{\text{f},q}},
    \end{equation} 
    
    \begin{equation}
        \mathbb{L}_x = \left(
\begin{array}{cccccccc}
 0 & 0 & 0 & 0 & 1 & 0 & 0 & 0 \\
 1 & 0 & 0 & 0 & 0 & 0 & 0 & -\frac{1}{c^2} \\
 0 & 0 & -\frac{\beta_{z \perp x}}{2} & \frac{\beta_{y \perp x}}{2} & 0 & -\frac{\beta_{z \perp x} \text{sgn}(a_x)}{2 \sqrt{\mu_0 \rho }} & \frac{\beta_{y \perp x} \text{sgn}(a_x)}{2 \sqrt{\mu_0 \rho }} & 0 \\
 0 & 0 & \frac{\beta_{z \perp x}}{2} & -\frac{\beta_{y \perp x}}{2} & 0 & -\frac{\beta_{z \perp x} \text{sgn}(a_x)}{2 \sqrt{\mu_0 \rho }} & \frac{\beta_{y \perp x} \text{sgn}(a_x)}{2 \sqrt{\mu_0 \rho }} & 0 \\
 0 & -\frac{\mathcal{C}_{\text{s},x}}{2 c^2} & -\frac{\mathcal{C}_{\text{f},x} \beta_{y \perp x} \text{sgn}(a_x)}{2 c^2} & -\frac{\mathcal{C}_{\text{f},x} \beta_{z \perp x} \text{sgn}(a_x)}{2 c^2} & 0 & -\frac{\alpha_{\text{f},x} \beta_{y \perp x}}{2 c \sqrt{\mu_0 \rho }} & -\frac{\alpha_{\text{f},x} \beta_{z \perp x}}{2 c \sqrt{\mu_0 \rho }} & \frac{\alpha_{\text{s},x}}{2 c^2 \rho } \\
 0 & \frac{\mathcal{C}_{\text{s},x}}{2 c^2} & \frac{\mathcal{C}_{\text{f},x} \beta_{y \perp x} \text{sgn}(a_x)}{2 c^2} & \frac{\mathcal{C}_{\text{f},x} \beta_{z \perp x} \text{sgn}(a_x)}{2 c^2} & 0 & -\frac{\alpha_{\text{f},x} \beta_{y \perp x}}{2 c \sqrt{\mu_0 \rho }} & -\frac{\alpha_{\text{f},x} \beta_{z \perp x}}{2 c \sqrt{\mu_0 \rho }} & \frac{\alpha_{\text{s},x}}{2 c^2 \rho } \\
 0 & -\frac{\mathcal{C}_{\text{f},x}}{2 c^2} & \frac{\mathcal{C}_{\text{s},x} \beta_{y \perp x} \text{sgn}(a_x)}{2 c^2} & \frac{\mathcal{C}_{\text{s},x} \beta_{z \perp x} \text{sgn}(a_x)}{2 c^2} & 0 & \frac{\alpha_{\text{s},x} \beta_{y \perp x}}{2 c \sqrt{\mu_0 \rho }} & \frac{\alpha_{\text{s},x} \beta_{z \perp x}}{2 c \sqrt{\mu_0 \rho }} & \frac{\alpha_{\text{f},x}}{2 c^2 \rho } \\
 0 & \frac{\mathcal{C}_{\text{f},x}}{2 c^2} & -\frac{\mathcal{C}_{\text{s},x} \beta_{y \perp x} \text{sgn}(a_x)}{2 c^2} & -\frac{\mathcal{C}_{\text{s},x} \beta_{z \perp x} \text{sgn}(a_x)}{2 c^2} & 0 & \frac{\alpha_{\text{s},x} \beta_{y \perp x}}{2 c \sqrt{\mu_0 \rho }} & \frac{\alpha_{\text{s},x} \beta_{z \perp x}}{2 c \sqrt{\mu_0 \rho }} & \frac{\alpha_{\text{f},x}}{2 c^2 \rho } \\
\end{array}
\right),
    \end{equation}

    \begin{equation}
        \mathbb{L}_y = \left(
\begin{array}{cccccccc}
 0 & 0 & 0 & 0 & 0 & 1 & 0 & 0 \\
 1 & 0 & 0 & 0 & 0 & 0 & 0 & -\frac{1}{c^2} \\
 0 & -\frac{\beta_{z \perp y}}{2} & 0 & \frac{\beta_{x \perp y}}{2} & -\frac{\beta_{z \perp y} \text{sgn}(a_y)}{2 \sqrt{\mu_0 \rho }} & 0 & \frac{\beta_{x \perp y} \text{sgn}(a_y)}{2 \sqrt{\mu_0 \rho }} & 0 \\
 0 & \frac{\beta_{z \perp y}}{2} & 0 & -\frac{\beta_{x \perp y}}{2} & -\frac{\beta_{z \perp y} \text{sgn}(a_y)}{2 \sqrt{\mu_0 \rho }} & 0 & \frac{\beta_{x \perp y} \text{sgn}(a_y)}{2 \sqrt{\mu_0 \rho }} & 0 \\
 0 & -\frac{\mathcal{C}_{\text{f},y} \beta_{x \perp y} \text{sgn}(a_y)}{2 c^2} & -\frac{\mathcal{C}_{\text{s},y}}{2 c^2} & -\frac{\mathcal{C}_{\text{f},y} \beta_{z \perp y} \text{sgn}(a_y)}{2 c^2} & -\frac{\alpha_{\text{f},y} \beta_{x \perp y}}{2 c \sqrt{\mu_0 \rho }} & 0 & -\frac{\alpha_{\text{f},y} \beta_{z \perp y}}{2 c \sqrt{\mu_0 \rho }} & \frac{\alpha_{\text{s},y}}{2 c^2 \rho } \\
 0 & \frac{\mathcal{C}_{\text{f},y} \beta_{x \perp y} \text{sgn}(a_y)}{2 c^2} & \frac{\mathcal{C}_{\text{s},y}}{2 c^2} & \frac{\mathcal{C}_{\text{f},y} \beta_{z \perp y} \text{sgn}(a_y)}{2 c^2} & -\frac{\alpha_{\text{f},y} \beta_{x \perp y}}{2 c \sqrt{\mu_0 \rho }} & 0 & -\frac{\alpha_{\text{f},y} \beta_{z \perp y}}{2 c \sqrt{\mu_0 \rho }} & \frac{\alpha_{\text{s},y}}{2 c^2 \rho } \\
 0 & \frac{\mathcal{C}_{\text{s},y} \beta_{x \perp y} \text{sgn}(a_y)}{2 c^2} & -\frac{\mathcal{C}_{\text{f},y}}{2 c^2} & \frac{\mathcal{C}_{\text{s},y} \beta_{z \perp y} \text{sgn}(a_y)}{2 c^2} & \frac{\alpha_{\text{s},y} \beta_{x \perp y}}{2 c \sqrt{\mu_0 \rho }} & 0 & \frac{\alpha_{\text{s},y} \beta_{z \perp y}}{2 c \sqrt{\mu_0 \rho }} & \frac{\alpha_{\text{f},y}}{2 c^2 \rho } \\
 0 & -\frac{\mathcal{C}_{\text{s},y} \beta_{x \perp y} \text{sgn}(a_y)}{2 c^2} & \frac{\mathcal{C}_{\text{f},y}}{2 c^2} & -\frac{\mathcal{C}_{\text{s},y} \beta_{z \perp y} \text{sgn}(a_y)}{2 c^2} & \frac{\alpha_{\text{s},y} \beta_{x \perp y}}{2 c \sqrt{\mu_0 \rho }} & 0 & \frac{\alpha_{\text{s},y} \beta_{z \perp y}}{2 c \sqrt{\mu_0 \rho }} & \frac{\alpha_{\text{f},y}}{2 c^2 \rho } \\
\end{array}
\right),
    \end{equation}

    \begin{equation}
        \mathbb{L}_z = \left(
\begin{array}{cccccccc}
 0 & 0 & 0 & 0 & 0 & 0 & 1 & 0 \\
 1 & 0 & 0 & 0 & 0 & 0 & 0 & -\frac{1}{c^2} \\
 0 & -\frac{\beta_{y \perp z}}{2} & \frac{\beta_{x \perp z}}{2} & 0 & -\frac{\beta_{y \perp z} \text{sgn}(a_z)}{2 \sqrt{\mu_0 \rho }} & \frac{\beta_{x \perp z} \text{sgn}(a_z)}{2 \sqrt{\mu_0 \rho }} & 0 & 0 \\
 0 & \frac{\beta_{y \perp z}}{2} & -\frac{\beta_{x \perp z}}{2} & 0 & -\frac{\beta_{y \perp z} \text{sgn}(a_z)}{2 \sqrt{\mu_0 \rho }} & \frac{\beta_{x \perp z} \text{sgn}(a_z)}{2 \sqrt{\mu_0 \rho }} & 0 & 0 \\
 0 & -\frac{\mathcal{C}_{\text{f},z} \beta_{x \perp z} \text{sgn}(a_z)}{2 c^2} & -\frac{\mathcal{C}_{\text{f},z} \beta_{y \perp z} \text{sgn}(a_z)}{2 c^2} & -\frac{\mathcal{C}_{\text{s},z}}{2 c^2} & -\frac{\alpha_{\text{f},z} \beta_{x \perp z}}{2 c \sqrt{\mu_0 \rho }} & -\frac{\alpha_{\text{f},z} \beta_{y \perp z}}{2 c \sqrt{\mu_0 \rho }} & 0 & \frac{\alpha_{\text{s},z}}{2 c^2 \rho } \\
 0 & \frac{\mathcal{C}_{\text{f},z} \beta_{x \perp z} \text{sgn}(a_z)}{2 c^2} & \frac{\mathcal{C}_{\text{f},z} \beta_{y \perp z} \text{sgn}(a_z)}{2 c^2} & \frac{\mathcal{C}_{\text{s},z}}{2 c^2} & -\frac{\alpha_{\text{f},z} \beta_{x \perp z}}{2 c \sqrt{\mu_0 \rho }} & -\frac{\alpha_{\text{f},z} \beta_{y \perp z}}{2 c \sqrt{\mu_0 \rho }} & 0 & \frac{\alpha_{\text{s},z}}{2 c^2 \rho } \\
 0 & \frac{\mathcal{C}_{\text{s},z} \beta_{x \perp z} \text{sgn}(a_z)}{2 c^2} & \frac{\mathcal{C}_{\text{s},z} \beta_{y \perp z} \text{sgn}(a_z)}{2 c^2} & -\frac{\mathcal{C}_{\text{f},z}}{2 c^2} & \frac{\alpha_{\text{s},z} \beta_{x \perp z}}{2 c \sqrt{\mu_0 \rho }} & \frac{\alpha_{\text{s},z} \beta_{y \perp z}}{2 c \sqrt{\mu_0 \rho }} & 0 & \frac{\alpha_{\text{f},z}}{2 c^2 \rho } \\
 0 & -\frac{\mathcal{C}_{\text{s},z} \beta_{x \perp z} \text{sgn}(a_z)}{2 c^2} & -\frac{\mathcal{C}_{\text{s},z} \beta_{y \perp z} \text{sgn}(a_z)}{2 c^2} & \frac{\mathcal{C}_{\text{f},z}}{2 c^2} & \frac{\alpha_{\text{s},z} \beta_{x \perp z}}{2 c \sqrt{\mu_0 \rho }} & \frac{\alpha_{\text{s},z} \beta_{y \perp z}}{2 c \sqrt{\mu_0 \rho }} & 0 & \frac{\alpha_{\text{f},z}}{2 c^2 \rho } \\
\end{array}
\right),
    \end{equation}
\end{subequations}
wherein $I_{8\times 8}$ is the 8D identity matrix, the dimensionless variables $\alpha_{\text{s/f},q}$ and $\beta_{q' \perp q}$ are given by \Autoref{eq:alphaBeta}, and
the new variables $\mathcal{C}_{\text{s/f},q} = \alpha_{\text{s/f},q} c_{\text{s/f},q}$ are the scaled slow/fast speeds. Note that the new eigenvalues slightly differ from those of Paper I in that it is the absolute value of the \alf{} speed that enters $\Lambda_q$. This is an important change as $c_{\text{s/f},q} \geq 0$, and we would like to treat the \alf{} mode on an equal footing to avoid unwanted switches between the reverse and forward \alf{} modes based on the value of $a_q$. 

The corresponding right eigenvectors can be computed according to $\mathbb{R}_q = \mathbb{L}_q^{-1}$, exploiting the following useful identities also found in \cite{roe1996}; $\alpha_{\text{s},q}^2 + \alpha_{\text{f},q}^2 = 1$, $\sum_{q' \neq q}\beta_{q'\perp q}^2 = 1$, $\alpha_{\text{s},q}^2 c_{\text{s},q}^2 + \alpha_{\text{f},q}^2 c_{\text{f},q}^2 = c^2$, $c_{\text{s},q} c_{\text{f},q} = s_q c$.

\bibliography{main.bbl}
\bibliographystyle{aasjournal}



\end{document}